\documentclass[11pt]{article}
\usepackage {graphicx}
\usepackage{amsmath,amsxtra,amssymb,latexsym, amscd}
\usepackage[mathscr]{eucal}

\makeindex
\begin{document}
\parindent=1.05cm 
\setlength{\baselineskip}{12truept} \setcounter{page}{1}
\makeatletter
\renewcommand{\@evenhead}{\@oddhead}
\renewcommand{\@oddfoot}{}
\renewcommand{\@evenfoot}{\@oddfoot}
\renewcommand{\thesection}{\arabic{section}.}
\renewcommand{\thesubsection}{\thesection\arabic{subsection}.}
\renewcommand{\theequation}{\thesection\arabic{equation}}
\@addtoreset{equation}{section}
\begin{center}
\vspace{10cm}
{\bf HIGH ENERGY SCATTERING OF DIRAC PARTICLES ON SMOOTH POTENTIALS}\\
\vspace{0.5cm}
\small{
Nguyen Suan Han$^{a}$\footnote{Email:lienbat76@gmail.com}, Le Anh Dung$^{a}$, \\
Nguyen Nhu Xuan$^{b}$, Vu Toan Thang$^{a}$\\
\vspace{0.5cm}
{$^a$\it Department of Theoretical Physics, Hanoi University of Science, Hanoi, Vietnam.} \\
{$^b$\it Department of Physics, Le Qui Don University, Hanoi, Vietnam.}\\}
\end{center}
\vspace{0.5cm}
\centerline{\bf Abstract}
\baselineskip=18pt
\bigskip
The derivation of the Glauber type representation for the high energy scattering
amplitude of particles of spin $1/2$ is given within the framework of the Dirac
equation in the Foldy $–$ Wouthuysen (FW) representation and two-component formalism.
The differential cross sections on the Yukawa and Gaussian potentials are also considered and discussed.
\vspace{0.5cm}
\newpage
\section{Introduction}
\indent Eikonal representation, or Glauber type representation, for the scattering amplitude of high energy particles with small scattering angles was first obtained in Quantum Mechanics \cite{1} and, then, in Quantum Field Theory based on the Logunov-Tavkhelize quasi-potential equation \cite{2},\cite{3}. The assumption of the smoothness of local pseudo-potential \cite{4}-\cite{6} allows us to explain successfully physical characteristics of high energy scattering of hadrons. More generally, it leads to a simple qualitative model of interactions between particles inthe asymptotic region of high energy.\\
\indent Eikonal representation for high energy scattering amplitude has been studied by other authors \cite{7}-\cite{19}. However, these investigations do not take into account the spin structure of the scattered particles. Some authors included spin effects in their studies \cite{18},\cite{19}, but their methods were not complete, or could not be applied for various potentials. On the other hand, experiments showed that spin effects, for example, the non-negligibility of the ratio of spin-flip to spin-nonflip amplitudes and Coulomb-hadron interference \cite{20a},\cite{21a}, play an important role in many physical processes, such as in the recent RHIC and LHC experiments \cite{20},\cite{21}. Moreover, present investigations did not utilize the FW transformation which is very convenient for passing to the quasi-classical approximation.Consequently, this paper aims to generalize the eikonal representation for the scattering amplitude of spinor particles, in particular, to establish the Glauber type representation for the scattering amplitude of spin $–1/2$ particles on smooth potentials at high energies within the framework of the Dirac equation in an external field after using the FW transformation \cite{22}-\cite{27}.\\
\indent The paper is organized as follows. In the second section, we obtain the Dirac equation in an external field in the FW representation. This representation has a special place in the field of relativistic quantum mechanics due to the following properties. First, Hamiltonian and quantum mechanical operators for relativistic particles in external fields in the FW representation are similar to those in the non-relativistic quantum mechanics. Second, the quasi-classical approximation and the classical limit of relativistic quantum mechanics can be obtained by a replacement of operators in quantum-mechanical Hamiltonian and equations of motion with the corresponding classical quantities \cite{24}-\cite{26}. This property is significant since most quantum effects are measured using classical apparatuses. Moreover, the FW representation is perfect for a description of spin effects which will be discussed later. We need also to mention that the relativistic FW transformation is widely and successfully used in quantum chemistry (see books \cite{30},\cite{31} and reviews \cite{33}-\cite{37}). In Section 3, employing the smoothness of external potentials and the Dirac equation in the FW representation, we end up with the Glauber type representation for the high energy scattering amplitude of spin $-1/2$ particles. In Section 4, the analytical expressions of the differential cross section in the Yukawa and Gaussian potentials are derived. The contribution of terms in the FW Hamiltonian to the scattering processes discussed. The results and possible generalizations of this approach are also discussed.
\section{Foldy – Wouthuysen transformation for the Dirac equation in external field}
\indent In general, there are two regular ways to perform the FW representation of the Dirac Hamiltonian: one approach gives a series of relativistic corrections to the nonrelativistic Schrodinger Hamiltonian, \cite{22},\cite{23},\cite{27}; the second approach allows one to obtain a compact expression for the relativistic FW Hamiltonian (see Refs. \cite{24}-\cite{26}, \cite{28}-\cite{29} and references therein). In this section, we utilize the method in Ref. \cite{29}.\\
 The Dirac equation for a particle with charge $e=q$ in an external electromagnetic field
 $V_\mu(V,\overrightarrow{A})$ is given by
\begin{equation}\label{e21}
i\frac{{\partial \Psi \left({\overrightarrow r ,t} \right)}}{{dt}} = {H_D}\Psi \left( {\overrightarrow r ,t} \right)
\end{equation}
with the Dirac hamiltonian $H_D$ and bi-spinor $\Psi$ defined as follows:
\begin{equation}\label{e22}
  {H_D} = \beta m + \mathcal{O} + \mathcal{E}
\end{equation}
\begin{equation}\label{e23}
\Psi=\begin{pmatrix}
             \psi \\
             \eta\\
           \end{pmatrix}
\end{equation}
where $\mathcal{O} = \overrightarrow \alpha  \left( {\overrightarrow p
- q\overrightarrow A } \right),\mathcal{E} = qV$, and $\psi,\eta$ are
two-component spinors. One can see that, the Dirac Hamiltonian (\ref{e23})
contains both the odd operator $\mathcal{O}$ and the even operator $\mathcal{E}$. The odd operator leads to the non-diagonal form of the Hamiltonian. As a result, the spinor $\Psi$ with positive energy is "mixed" with the negative-energy one $\eta$. However, it is necessary to isolate the positive-energy (particle) spinor, which will be employed in the next section to derive the scattering amplitude. Let us consider the following unitary transformation \cite{29}
\begin{equation}\label{e24}
\Psi _{FW} = U\Psi,\quad U = \frac{{1 + \sqrt {1 + {X^2}}  + \beta X}}{{\sqrt {2\sqrt {1 + {X^2}} \left( {1 + \sqrt {1 + {X^2}} } \right)} }},{\rm{  }}X = \frac{\mathcal{O}}{m}.
\end{equation}
with (\ref{e24}), the Dirac Hamiltonian is transformed as
\begin{equation}\label{e25}
H_{FW} = i{\partial _t} + U\left( {{H_D} - i{\partial _t}} \right){U^{ - 1}}.
\end{equation}
The explicit expression for the FW Hamiltonian is \cite{29}
\begin{equation}\label{e26}
{H_{FW}} = \beta \varepsilon  + {\mathcal E} - \frac{1}{8}{\left\{ {\frac{1}{{\varepsilon \left( {\varepsilon  + m} \right)}},\left[ {{\mathcal O},\left[ {{\mathcal O},{\mathcal F}} \right]} \right]} \right\}_ + },
\end{equation}
where
\begin{equation}\label{e27}
  \varepsilon  = \sqrt {{m^2} + {{\mathcal O}^2}} ,\quad {\mathcal F} = {\cal E} - i{\partial _t}.
\end{equation}
In Eq. (\ref{e26}), commutators of the third and higher orders as well as degrees of commutators of the third and higher orders are disregarded.\\
\indent In the specific case, when the external field is scalar $(\vec{A}=0)$, eq.(\ref{e26}) becomes
\begin{equation}\label{e28}
H_{FW} = \beta \varepsilon  + qV + \frac{q}{8}\left\{ {\frac{1}{{\varepsilon \left( {\varepsilon  + m} \right)}},\overrightarrow {i\Sigma } .\left( {p \times \nabla V} \right) + 2\overrightarrow \Sigma  .\left( {\nabla V \times p} \right) + {\nabla ^2}V} \right\}_ +
\end{equation}
In our present study, we obtain explicit relations for the scattering of a nonrelativistic particle, while we plan to consider the corresponding relativistic problem in the future.\\
\indent Using the non-relativistic approximation, $\varepsilon=\sqrt{m^2 + {\overrightarrow p ^2}}\approx m + \frac{\overrightarrow p ^2}{2m}$, the FW  Hamiltonian (\ref{e28}) to the order $\left(\frac{1}{m^2}\right)$ can be written as
\begin{equation}\label{e29}
  H_{FW}= \beta \left( m + \frac{\overrightarrow p^2}{2m} \right) + qV + \frac{iq}{8m^2}\overrightarrow\Sigma.\left(p \times \nabla V\right) + \frac{q}{4m^2}\overrightarrow\Sigma.\left(\overrightarrow\nabla V \times \overrightarrow p\right) + \frac{q}{8m^2}\nabla ^2V.
\end{equation}
In our study, the external field is a scalar central potential $V=V(r)$. The FW Hamiltonian, therefore, becomes
\begin{equation}\label{e210}
  H_{FW} = \beta \left(m + \frac{\overrightarrow p ^2}{2m}\right) + qV + \frac{q}{4m^2r}\frac{dV}{dr}\overrightarrow\Sigma.\overrightarrow L  + \frac{q}{8m^2}\nabla ^2V.
\end{equation}
Due to the $\beta-$ matrix, the FW Hamiltonian (\ref{e210}) contains relativistic corrections for both particle and anti-particle which include the spin – orbit coupling and the Darwin term. The Darwin term is added to direct interaction of charged particles as point charges, and it characterizes the Zitterbewegung motion of Dirac particles. It is related to particles in the FW representation being not concentrated atone point but rather spreading out over a volume with radius of about $\left(\frac{1}{m}\right)$ \cite{23}.\\
\indent Since the only particle is considered in our scattering problem, one needs to deal with the positive – energy component of the Hamiltonian (\ref{e210})
\begin{equation}\label{e211}
  H_{FW}^+ = m + \frac{\overrightarrow p^2}{2m} + qV + \frac{q}{4m^2r}\frac{dV}{dr}\overrightarrow\sigma.\overrightarrow L  + \frac{q}{8m^2}\nabla ^2V.
\end{equation}
It is also important to note that, the relativistic correction terms guarantee that the wave function in the FW representation agrees with the non-relativistic Pauli wave function for spin $– 1/2$ particles \cite{23}. This Hamiltonian (\ref{e211}) include the contribution of the Darwin term in the scattering amplitude. As shown in the next section, this term leads to different result compare with that obtained in Reference \cite{17}.
\section{Glauber type representation for scattering amplitude}
\indent With Hamiltonian $H_{FW}^+$, the equation for two-component wave function $\psi(\overrightarrow r ,t)$ is given by
\begin{equation}\label{e31}
  \left[m + \frac{\overrightarrow p ^2}{2m} + eV + \frac{e}{4m^2r}\frac{dV}{dr}\overrightarrow\sigma.\overrightarrow L  + \frac{e}{8m^2}\nabla ^2V \right]\psi(\overrightarrow r ,t)= i\frac{\partial \psi(\overrightarrow r ,t)}{\partial t}.
\end{equation}
By the variable separation
\begin{equation}\label{e32}
  \psi(\overrightarrow r ,t) = e^{- iEt}\varphi(\overrightarrow r),
\end{equation}
Equation (\ref{e31}) can be reduced to
\begin{equation}\label{e33}
  \left[m - \frac{\overrightarrow\nabla^2}{2m^2}\rm{+ e}V - \frac{ie}{4m^2r}\frac{dV}{dr}\left(\overrightarrow\sigma\times\overrightarrow r \right)\overrightarrow\nabla+ \frac{e}{8m^2}\nabla ^2V - E\right]\varphi(r) = 0.
\end{equation}
The solution to (\ref{e33}) will be sought in the form
\begin{equation}\label{e34}
\varphi(r) = e^{ipz}\varphi^{( + )}(r) + e^{- ipz}\varphi ^{( - )}(r) = e^{ipz}\varphi^{( + )}(b,z) + e^{- ipz}\varphi ^{( - )}(b,z),
\end{equation}
where $\overrightarrow r=(\overrightarrow b ,z)$, and the z-axis is chosen to be coincident with the direction of incident momentum $\overrightarrow{p}$. Two – component spinors $\varphi^{(+)}(r)$ and $\varphi^{(-)}(r)$ satisfy the following boundary conditions
\begin{equation}\label{e35}
  \varphi ^{(+)}(\overrightarrow b ,z)|_{z \to  - \infty} = \varphi _0,\varphi^{(-)}(\overrightarrow b ,z)|_{z \to  - \infty } = 0.
\end{equation}
Two terms in (\ref{e34}) describe the propagation of incident and reflected waves along the $z$-axis, respectively. Substitution of (\ref{e34}) into (\ref{e33}) yields
\begin{equation}\label{e36}
\begin{split}
\frac{e^{ipz}}{2m}&\Biggr[
\left(U + \frac{1}{8m^2}\nabla ^2U + \frac{1}{4m^2r}\frac{dU}{dr}p\left(\overrightarrow \sigma\times \overrightarrow r\right)_z- 2ip\frac{\partial }{\partial z}\right)\varphi^{(+)}\\
 &-\nabla^2\varphi^{(+)} - \frac{i}{4m^2r}\frac{dU}{dr}\left(\overrightarrow \sigma   \times \overrightarrow r\right)\nabla \varphi^{(+)}
\Biggr]\\
+\frac{e^{- ipz}}{2m}&\Biggr[
\left(U + \frac{1}{8m^2}\nabla ^2U - \frac{1}{4m^2r}\frac{dU}{dr}p\left(\overrightarrow \sigma   \times \overrightarrow r\right)_z + 2ip\frac{\partial}{\partial z} \right)\varphi^{(-)}\\
&-\nabla ^2\varphi ^{(-)} - \frac{i}{4m^2r}\frac{dU}{dr}\left(\overrightarrow \sigma   \times \overrightarrow r\right)\nabla\varphi^{(-)}
\Biggr] = 0,
\end{split}
\end{equation}
where $U(\overrightarrow r) = 2meV(\overrightarrow r)$ and $E\approx m + \frac{\overrightarrow p^2}{2m}$ (in the non-relativistic approximation). Due to the smoothness of the potential , the quasi-classical condition of scattering is satisfied \cite{17},\cite{30}
\begin{equation}\label{e37}
  \left|\frac{\dot U}{Up}\right| = \left|\frac{\dot V}{Vp} \right| \ll 1,\left|\frac{U}{p^2}\right| = \left| \frac{2meV}{p^2}\right| \ll 1.
\end{equation}
With the condition (\ref{e37}), the spinors $\varphi^{\pm}({b,z})$ are slowly varying functions and approximately satisfy the equations
\begin{equation}\label{e38}
  \left\{\begin{array}{l}
\left(U + \frac{1}{8m^2}\nabla ^2U + \frac{1}{4m^2r}\frac{dU}{dr}p(\overrightarrow\sigma\times \overrightarrow r)_z\right)\varphi^{(+)}(\overrightarrow b,z) = 2ip\frac{\partial \varphi ^{(+)}(\overrightarrow b ,z)}{\partial z}\\
\left(U + \frac{1}{8m^2}\nabla^2U - \frac{1}{4m^2r}\frac{dU}{dr}p(\overrightarrow\sigma\times \overrightarrow r)_z\right)\varphi^{(-)}( \overrightarrow b ,z) =  - 2ip\frac{\partial \varphi ^{(-)}(\overrightarrow b ,z)}{\partial z}
\end{array} \right.
\end{equation}
Note that
\begin{equation}\label{e39}
\left(\overrightarrow\sigma\times\overrightarrow r\right)_z = \left(\overrightarrow\sigma\times\overrightarrow b\right)_z + \left(\overrightarrow\sigma\times\overrightarrow z\right)_z = \left(\overrightarrow\sigma\times \overrightarrow b\right)_z =- b\left( \frac{\overrightarrow b}{b}\times\overrightarrow\sigma\right)_z= - b\left(\overrightarrow n \times\overrightarrow\sigma\right)_z,
\end{equation}
here $\overrightarrow n = \frac{\overrightarrow b}{b} =(\sin \phi ,\cos \phi)$ where $\phi$ is the azimuthal angle in the $(x,y)$ - plane.  Equations in (\ref{e38}) can be rewritten as
\begin{equation}\label{e310}
\left\{ \begin{array}{l}
\left( {U + \frac{1}{{8{m^2}}}{\nabla ^2}U - \frac{{pb}}{{4{m^2}r}}\frac{{dU}}{{dr}}{{\left( {\overrightarrow n  \times \overrightarrow \sigma  } \right)}_z}} \right){\varphi ^{\left(  +  \right)}}\left( {\overrightarrow b ,z} \right) = 2ip\frac{{\partial {\varphi ^{\left(  +  \right)}}\left( {\overrightarrow b ,z} \right)}}{{\partial z}}\\
\left( {U + \frac{1}{{8{m^2}}}{\nabla ^2}U + \frac{{pb}}{{4{m^2}r}}\frac{{dU}}{{dr}}{{\left( {\overrightarrow n  \times \overrightarrow \sigma  } \right)}_z}} \right){\varphi ^{\left(  -  \right)}}\left( {\overrightarrow b ,z} \right) =  - 2ip\frac{{\partial {\varphi ^{\left(  -  \right)}}\left( {\overrightarrow b ,z} \right)}}{{\partial z}}
\end{array} \right.
\end{equation}
The solutions of equations (\ref{e310}) with the boundary conditions (\ref{e35}) can be written in the form
\begin{align}
\varphi^{(+)}(\overrightarrow b ,z)=&\varphi _0\exp\left\{ \frac{1}{2ip}\int\limits_{- \infty}^z \left[U(r) + \frac{1}{8m^2}\left(\nabla ^2U(r) \right)- \frac{pb}{4m^2r}\frac{dU}{dr}\left(\overrightarrow n\times\overrightarrow\sigma\right)_z\right]dz'\right\},\label{e311} \\
\varphi^{(-)}(\overrightarrow b ,z)=& 0.\label{e312}
\end{align}
From the boundary condition (\ref{e35}), one can see that the reflected wave equals to zero. From (\ref{e34}), the wave function for scattering particle is
\begin{equation}\label{e313}
\varphi(r) = e^{ipz}\varphi_0.\exp \left[\chi _0(b,z)+ i\left(\vec{n} \times\vec{\sigma}\right)_z\chi _1(b,z)\right],
\end{equation}
where functions $\chi _0(b,z)$ and $\chi _1(b,z)$ are defined as
\begin{align}
  \chi _0(\overrightarrow b,z) =&\frac{1}{2ip}\int\limits_{-\infty}^z \left[U(r)+\frac{1}{8m^2}\left(\nabla^2U(r)\right)\right]dz' \label{e314}\\
  \chi_1(\overrightarrow b,z)=&\frac{b}{8m^2}\int\limits_{-\infty}^z\frac{1}{r}\frac{dU}{dr}dz'\label{e315}
\end{align}
For the scattering amplitude, we obtain respectively
\begin{equation}\label{e316}
  \begin{split}
f(\theta) =&-\frac{1}{4\pi}\mathop \int d\mathbf{r}e^{- i\mathbf{p}'\mathbf{r}}\varphi_0^*(\mathbf{p}')\left[U + \frac{\nabla^2U}{8m^2} - \frac{pb}{4m^2r}\frac{dU}{dr}
(\overrightarrow n\times\overrightarrow\sigma)_z\right]\varphi(r)\\
 =&\frac{p}{2i\pi}\mathop\int d^2\mathbf{b}e^{-ib\bf{\Delta}}\varphi_0^*(\mathbf{p}')\left[e^{\chi _0+ i\left(n\times\sigma\right)_z\chi _1} - 1\right]\varphi _0(\mathbf{p}).\\
  \end{split}
\end{equation}
One can rewrite this formula as
\begin{equation}\label{e317}
f(\theta) = \varphi _0^*(\vec p')\left[{A( \theta) + {\sigma _y}B( \theta)} \right]{\varphi _0}(\vec p)
\end{equation}
here \cite{38}
\begin{equation}\label{e318}
\varphi_0 = \left(\begin{array}{l}
1\\
0
\end{array}\right) \hspace{6pt}\text{or}\hspace{6pt}\varphi_0 = \left(\begin{array}{l}
0\\
1
\end{array}\right)\hspace{6pt}\text{for}\hspace{6pt} \lambda = \frac{1}{2} \hspace{6pt}\text{or} \hspace{6pt}\lambda = -\frac{1}{2}
\end{equation}
\begin{align}
\Delta =&\vec p'- \vec p = 2p\sin \frac{\theta}{2}; \hspace{6pt} \chi_0= \chi _0(\vec b,\infty),\chi_1= \chi_1(\vec b,\infty) \label{e319}\\
A(\theta)=&- ip\int\limits_0^\infty {bdb}J_0(b\Delta)\left[e^{\chi _0}\cos{\chi_1}- 1\right]\label{e320}\\
B(\theta)=&- ip\int\limits_0^\infty {bdb}J_1(b\Delta)e^{\chi _0}\sin{\chi_1}\label{e321}
\end{align}
where $p'$ and $\theta$ are the momentum after scattering and the scattering angle; $J_0(b\Delta)$ and $J_1(b\Delta)$ are the Bessel functions of the zezoth and the first order. The presence of quantities $A(\theta)$ and $B(\theta)$ determined by formulas (\ref{e320}) and (\ref{e321}) in the high – energy limit shows that there are both spin-flip and non spin-flip parts contributing to the scattering amplitude.
\section{Differential scattering cross section}
\indent In this section, using the obtained expression for the scattering amplitude shown above, we derive the differential cross sections for the scattering in Yukawa and Gaussian potentials for cases in which the Darwin term is included or excluded, respectively. In fact, the Gaussian potential is a smooth and non-singular function that ensures the constancy of the total hadron cross section \cite{39a}. Those will then be used to evaluate the contribution of the Darwin term in different regions of momentum and to study the behaviour of the Coulomb – nuclear interference in our in-process studies \cite{44a}.
In this section, using the obtained expression for the scattering amplitude above, we derive the differential cross sections for the scattering in Yukawa and Gaussian potentials for cases in which the Darwin term is included or excluded, respectively. Those will then be used to evaluate the contribution of the Darwin term in different region of momentum.
\subsection{Yukawa potential}
Let us consider the Yukawa potential \cite{39} given by
\begin{equation}\label{e41}
  U(r)= \frac{g}{r}e^{-\mu r}= \frac{g}{r}e^{- \frac{r}{R}}
\end{equation}
here, $g$ is a magnitude scaling constant whose dimension is of energy, $\mu$ is another scaling constant which is related to R - the effective size where the potential is non-zero – as $\mu=\frac{1}{R}$.\\
\indent From (\ref{e314}) and (\ref{e315}), one gets (see Appendix A)
\begin{align}
\chi_0(b)=&\frac{{\pi g}}{{ip}}\left( {1 + \frac{{{\mu ^2}}}{{8{m^2}}}} \right){K_0}\left( {\mu b} \right),\label{e42}\\
\chi_1(b)=& - \frac{\mu \pi g}{4m^2}K_1(\mu b),\label{e43}
\end{align}
where $K_0(\mu b)$ and $K_1(\mu b)$ is the MacDonald function of zeroth order \cite{45} and first order, respectively. Substitution of (\ref{e42}) and (\ref{e43}) into (\ref{e320}) and (\ref{e321}) yields
\begin{align}
A(\theta)=&-\frac{\pi g\left(1 + \frac{\mu ^2}{8m^2}\right)}{\Delta ^2 + {\mu ^2}},\label{e44}\\
B(\theta)=&\frac{i\pi gp}{4m^2}.\frac{\Delta}{\mu ^2+ \Delta ^2}.\label{e45}
\end{align}
The differential cross section is then
\begin{equation}\label{e46}
\frac{d\sigma}{d\Omega}\left|_{Y_D}\right. = \left| {A\left( \theta  \right)} \right|^2 + \left| {B\left( \theta  \right)} \right|^2 =
 \frac{\pi ^2g^2}{\left[4p^2\sin^2\left(\theta/2\right)+ \mu^2\right]^2}\left[\left(1 + \frac{\mu ^2}{8m^2}\right)^2 + \frac{p^4\sin ^2\left( \theta/2\right)}{4m^4}\right].
\end{equation}
This expression of differential cross section is for the case in which the Darwin term is included. If we ignore this term, the differential cross section is
\begin{equation}\label{e47}
\frac{d\sigma}{d\Omega}\left|_{Y_o}\right. =\frac{\pi ^2g^2}{\left[\mu ^2+ 4p^2\sin^2(\theta /2)\right]^2}\left[ 1 + \frac{p^4\sin ^2(\theta/2)}{4m^4}\right].
\end{equation}
With a dimensionless q defined as $q = \frac{p}{\mu}$ \cite{40}, one can rewrite expressions (\ref{e46}) and (\ref{e47}), respectively, as
\begin{align}
\frac{{d\sigma}}{{d\Omega }}\left| {_{{Y_D}}} \right. = &\frac{\pi ^2g^2}{\mu ^4\left[4q^2\sin^2(\theta/2)+ 1\right]^2}\left[\left(1 + \frac{1}{8q^2} \right)^2+ \frac{\mu ^4q^4}{4m^4}\sin^2(\theta/2)\right]\label{e48},\\
\frac{d\sigma}{d\Omega}\left|_{Y_o}\right. =&\frac{\pi ^2g^2}{\mu ^4}{\left[1 + 4q^2\sin^2(\theta /2)\right]^2}\left[1 + \frac{\mu^4q^4}{4m^4}\sin ^2(\theta /2)\right].
  \label{e49}
\end{align}
The dependence of the differential cross section on q (or, in other words, on the incident momentum) and the scattering angle $\theta$ in both two cases are graphically illustrated in Figs. 4.1 and 4.2 (constants are set to unit).
\begin{center}
\vspace{-10pt}
\includegraphics[width=11cm]{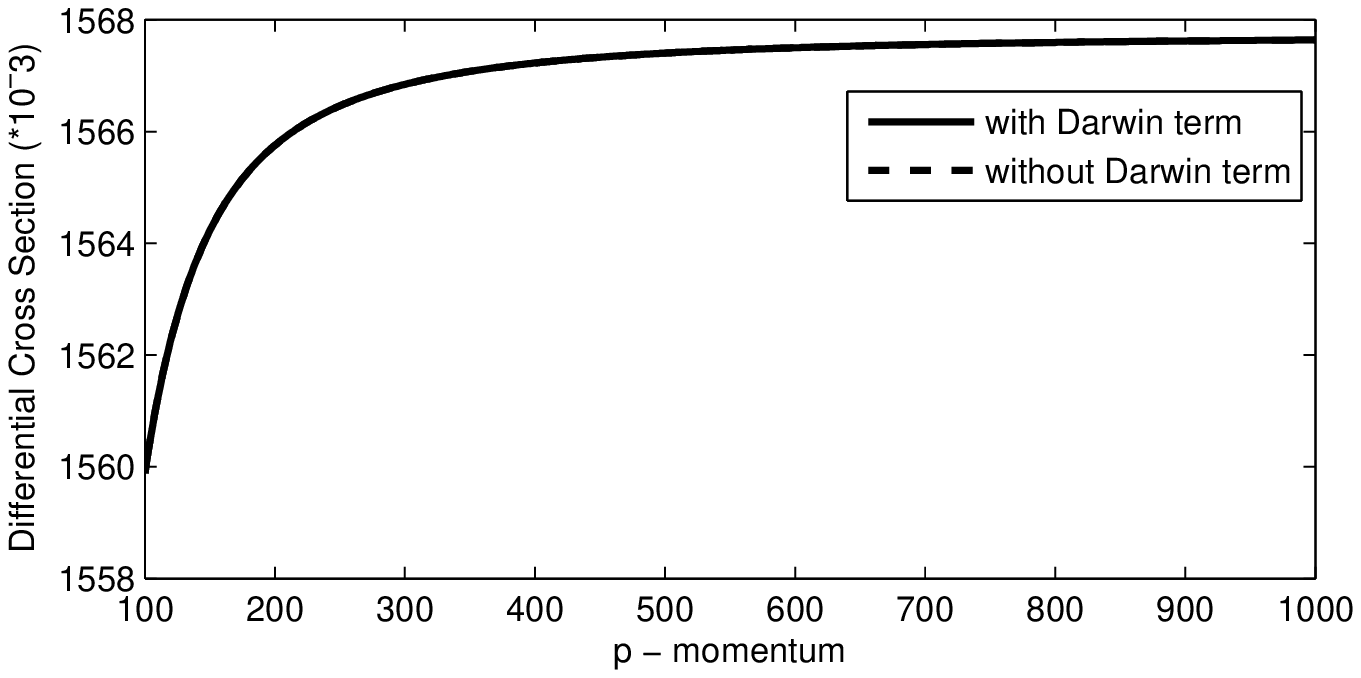}\\
\vspace{-6pt}(a)\\
\includegraphics[width=11cm]{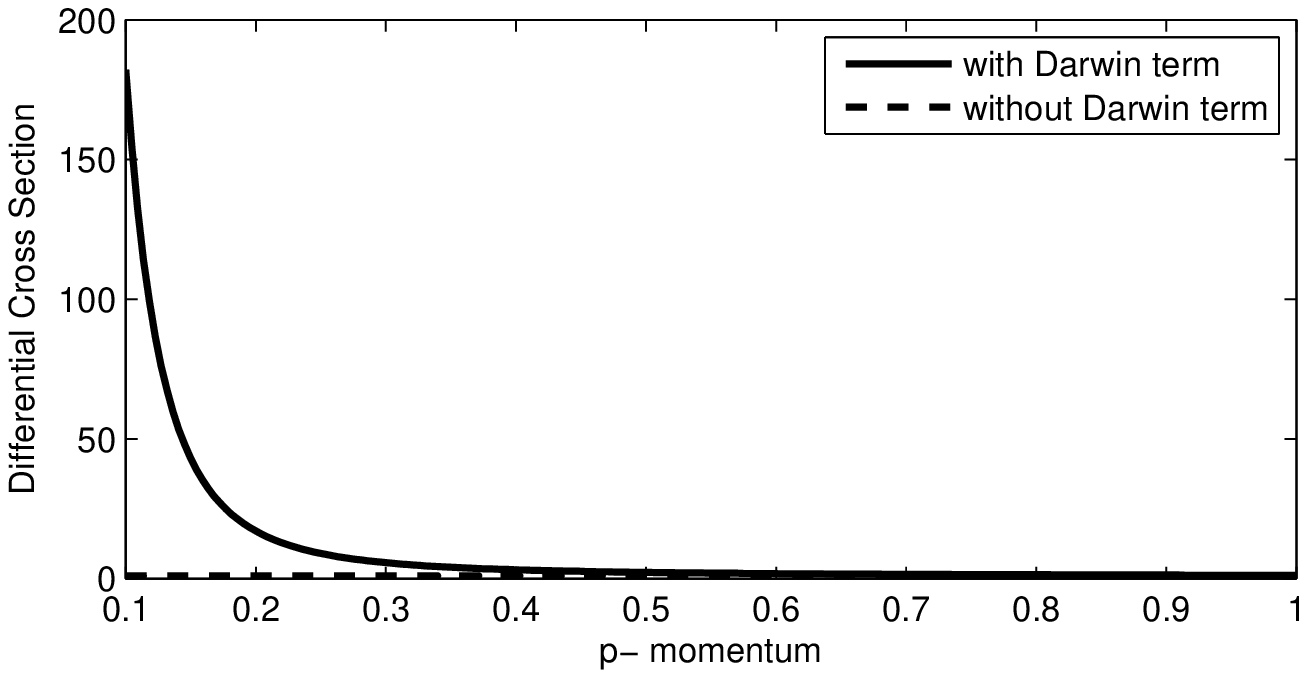}\\
(b)\\
\small{\textbf{Fig.4.1.} Dependence of the differential cross section on the momentum of incident\\ particle (with a specific small value of the scattering angle, $\theta=0.1 rad$)}\\
(a) for large p - momentum.   \hspace{2cm}         (b) for small p - momentum.
\end{center}
\begin{center}
\vspace{-10pt}
\includegraphics[width=7cm]{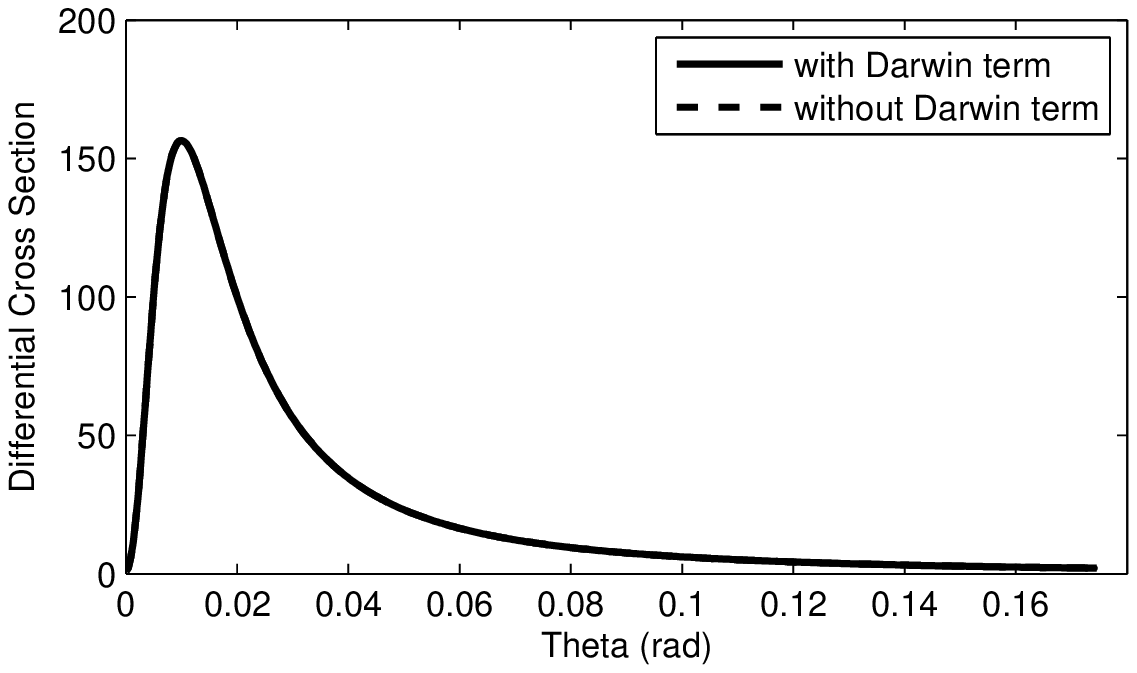}\includegraphics[width=7cm]{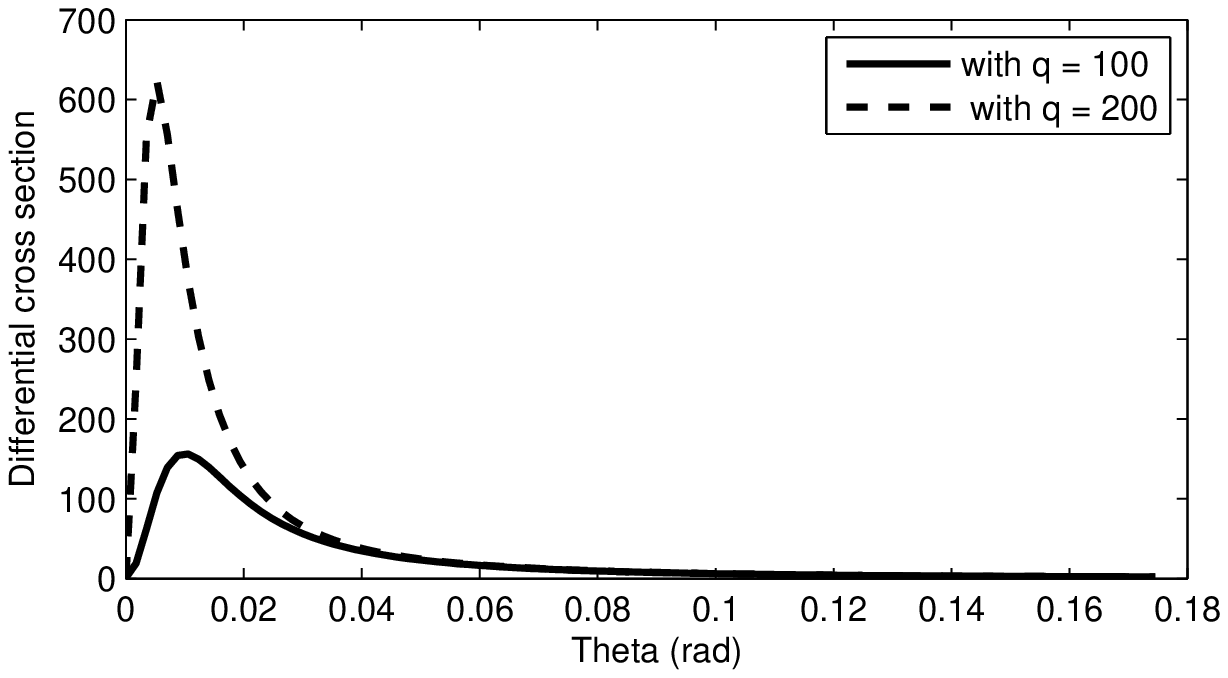}\\
\hspace{40pt} (a)     \hspace{200pt}  (b)\\
\small{\textbf{Fig.4.2.} Dependence of the differential cross section on the scattering angle:\\
 (a) Differential cross section with and without the Darwin term with $q=100$.\\
 (b) Differential cross section with the Darwin term with $q=100$ and $q=200$.}
\end{center}
In Fig.4.2, the differential cross section has a peak at a small value of scattering angle. Also, the behavior of the differential cross section in those  figures is similar to one  obtained formerly in references \cite{39}-\cite{41}.
\subsection{Gaussian Potential}
Now, we consider the Gaussian potential of following form \cite{39}
\begin{equation}\label{e410}
U\left( r \right) = \lambda {e^{ - \alpha {r^2}}} = \lambda \exp \left( { - \frac{{{r^2}}}{{2{R^2}}}} \right),
\end{equation}
where $\lambda$ is a magnitude scaling constant, $R$ is the effective size where the potential is non-zero and $\alpha$ is another scaling constant, $\alpha=\frac{1}{2R^2}$.\\
\indent To get the differential cross section, we performed some calculations similar to the calculation of the differential cross section with Yukawa potential in subsection 4.1 above (see Appendix B for detail). As a result, we obtain
\begin{equation}\label{e411}
\frac{{d\sigma }}{{d\Omega }}\left| {_{{G_D}}} \right. =\frac{{\pi {\lambda ^2}}}{{16{\alpha ^3}}}\exp \left( { - \frac{{2{p^2}{{\sin }^2}(\theta /2)}}{\alpha }} \right)\left[ {{{\left( {1 - \frac{{{p^2}}}{{2{m^2}}}{{\sin }^2}(\theta /2)} \right)}^2} + \frac{{{p^4}}}{{4{m^4}}}{{\sin }^2}(\theta /2)} \right].
\end{equation}
Now, if the Darwin term is ignored, the differential cross section is
\begin{equation}\label{e412}
\frac{{d\sigma }}{{d\Omega }}\left| {_{{G_o}}} \right. = \frac{{\pi {\lambda ^2}}}{{16{\alpha ^3}}}\exp \left( { - \frac{{2{p^2}{{\sin }^2}(\theta /2)}}{\alpha }} \right)\left( {1 + \frac{{{p^4}{{\sin }^2}(\theta /2)}}{{4{m^4}}}} \right).
\end{equation}
Figs.$4.3$ – $4.4$ graphically describe the dependence of the differential cross section on the momentum of incident particle and the scattering angle (constants are set to unit).
\begin{center}
\vspace{-8pt}
\includegraphics[width=10cm]{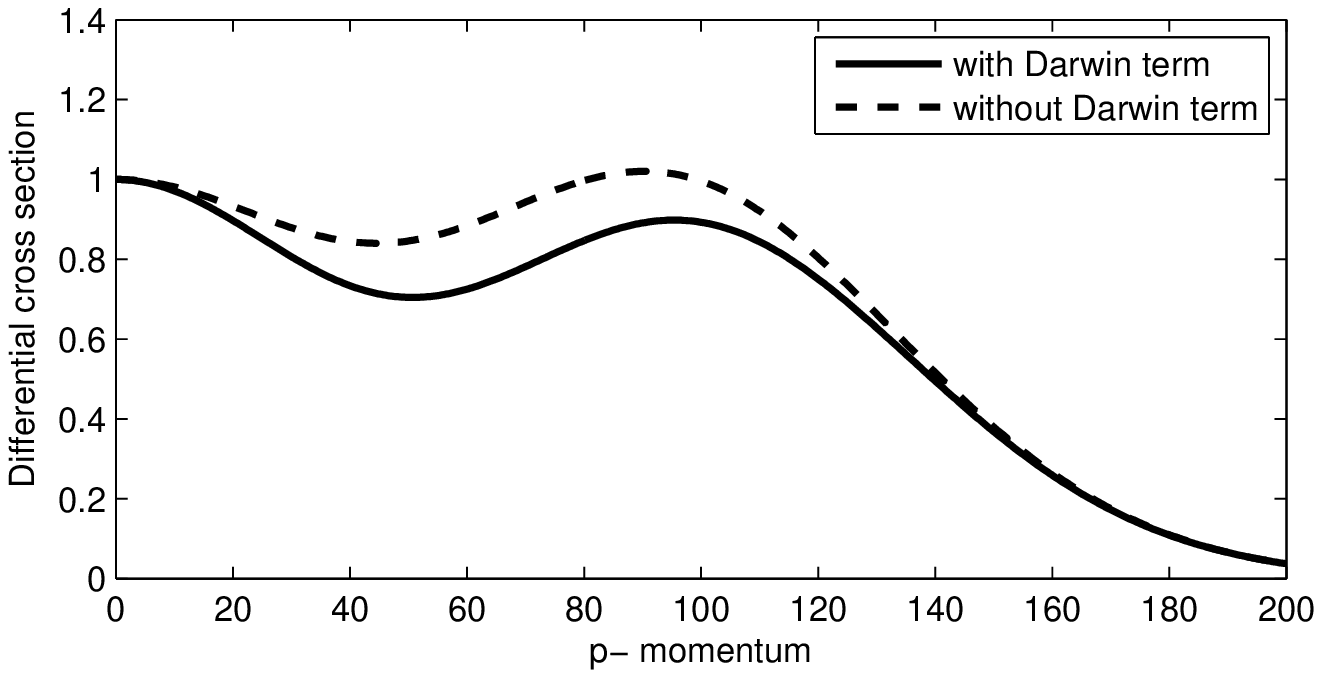}\\
\vspace{-3pt}\small{(a)}\\
\includegraphics[width=10cm]{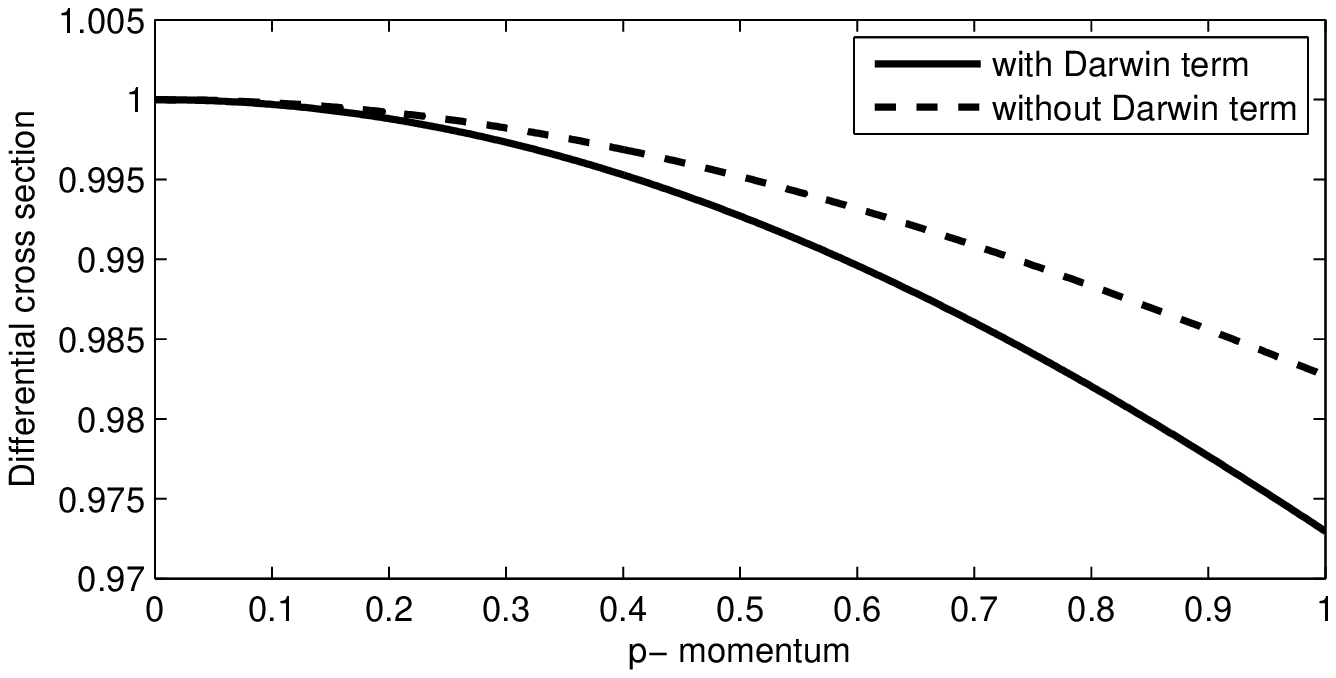}\\
\small{(b)}\\
\small{\textbf{Fig.4.3.} Dependence of the differential cross section on the momentum of incident particle (at a particular small value of the scattering angle)\\
 (a) with  large p – momentum. \hspace{50pt}   (b) with small p - momentum.}
\end{center}
\begin{center}
\includegraphics[width=11cm]{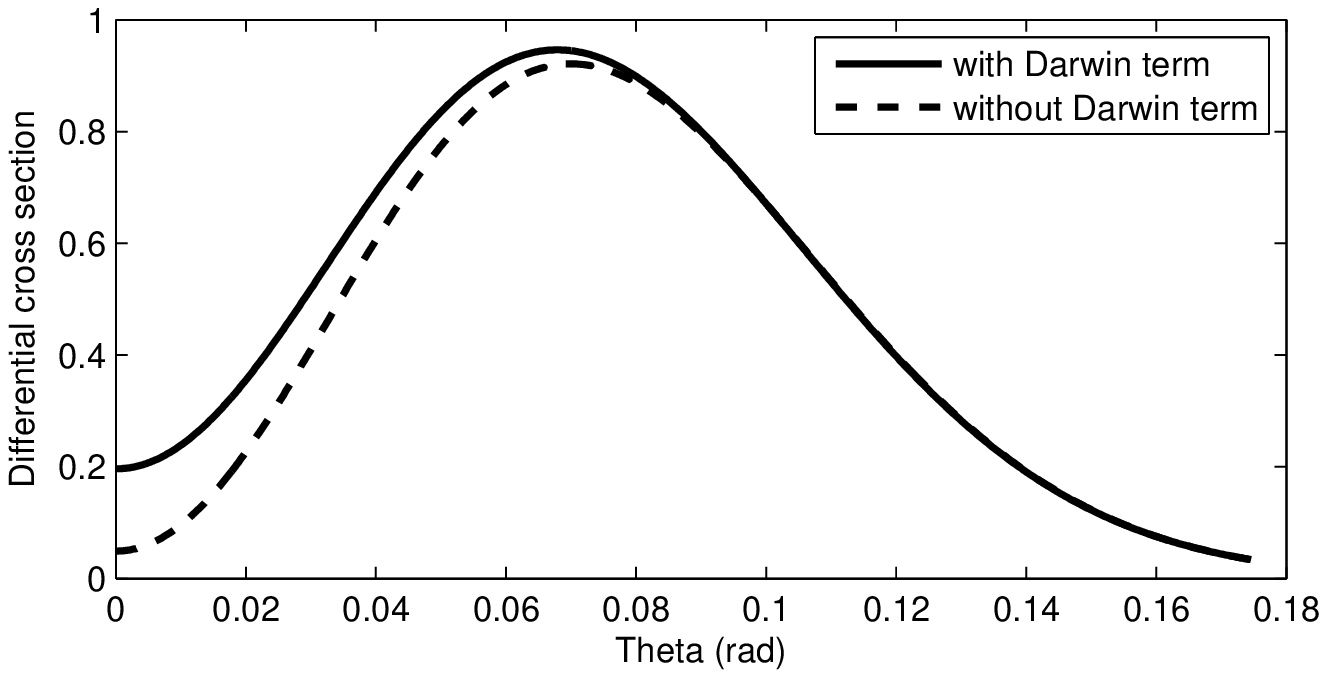}\\
(a)\\
\includegraphics[width=11cm]{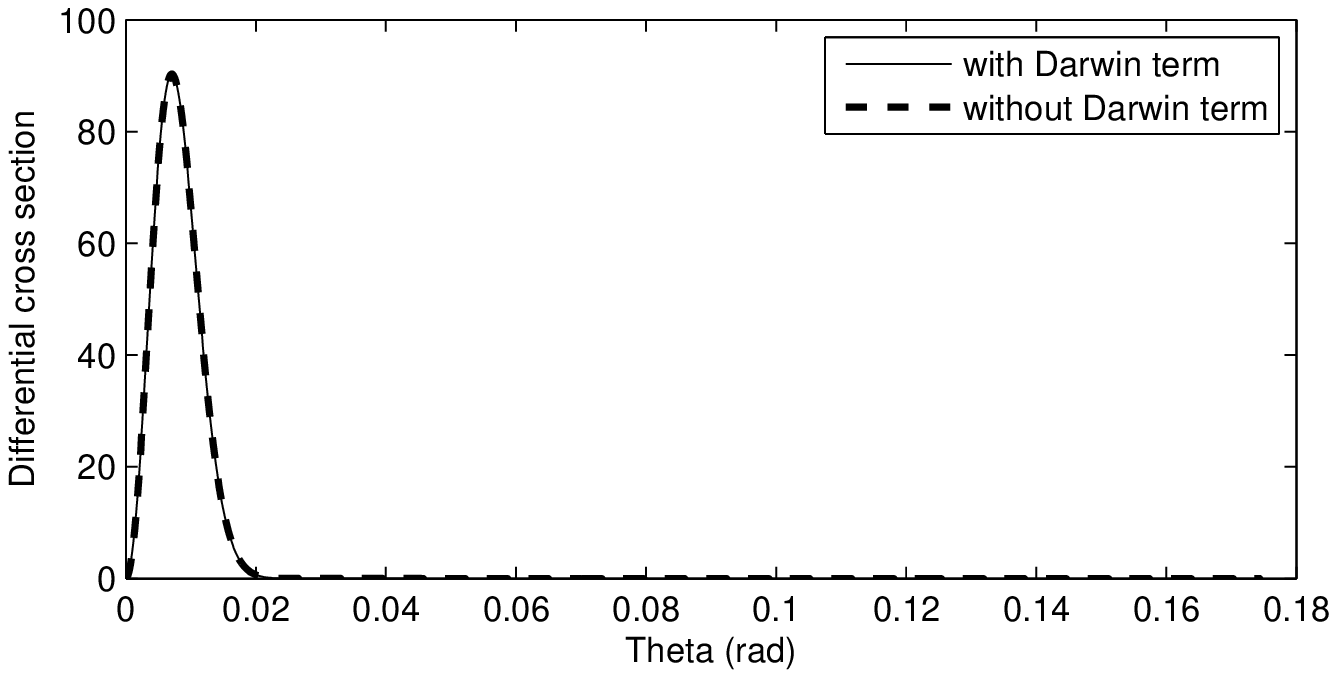}\\
(b)\\
\small{\textbf{Fig.4.4.} Dependence of the differential cross section on the scattering angle:\\
a) at $p=100$  \hspace{3cm} b) at $p=100$}
\end{center}
Unlike the case of Yukawa potential considered above, in the case of Gaussian potential the Darwin term causes non-negligible contributions to the differential cross section as shown in Figs.4.3 and 4.4. In the region of small values of momentum and very small scattering angles, the contribution of the Darwin term is significant.
\section{Conclusion}
By employing the step-by-step FW transformation which consists of two unitary transformation to the order $\left(\frac{1}{m^2}\right)$, we obtained the non-relativistic expression for Dirac Hamiltonian in the FW representation, which describes the interaction of particles and anti-particles having spin $1/2$ with an electromagnetic field. With the assumption of smooth potential, we ended up with the Glauber type representation for the high energy scattering amplitude of Dirac particles with small scattering angles. The resultant scattering amplitude includes the contribution of the Darwin term. This term guarantees that the wave function in the FW representation agrees with the non-relativistic Pauli wave function for spin $1/2$   particles. The expressions for the differential cross section with and without the Darwin term in the Yukawa and Gaussian potentials, which are two different forms of nuclear potential serving for the problem of Coulomb – nuclear interference \cite{44a}, are derived, respectively. We showed that the Darwin term has relatively significant contribution at some finite ranges of incident particle’s momentum $p$. However, this contribution is very small for large particle’s momenta. For the problem of scattering on the gravitational field, due to the relation to some basic problems such as strong gravitational forces near black-holes, string modification of theory of gravity and some other effects of quantum gravity \cite{13}, \cite{42}-\cite{44a}, the Darwin term derived in Ref.\cite{46} might be important; and, therefore, in our point of view, the application and generalization of the method proposed in this paper are necessary.
\section*{Acknowledgments}
We are  grateful to Profs. B.M. Barbashov,  A.V. Efremov, M.K.Volkov, O. V. Selyugin and V.V.Nesterenko for useful discussions. N.S.H. is indebted to Profs. V.N. Pervushin, A,J. Silenko  for reading the manuscript and making useful remarks for improvements, N.S.H. is also indebted to Prof. A. B. Arbuzov for support during his stay at JINR in Dubna and warm hospitality.This research is funded by NAFOSTED under grant number 103.03-2012.02, by the Joint Institute for Nuclear Research Dubna.
\appendix
\renewcommand{\thesection}{\Alph{section}.}
\section{{Appendix A. Yukawa Potential}}
For the Yukawa potential (\ref{e41}), since
\begin{equation}\label{A1}
  \frac{1}{r}\frac{dU}{dr} =\frac{g}{r}\frac{d}{dr}\left({\frac{{{e^{ - \mu r}}}}{r}} \right) = \frac{{ - g\left( {1 + \mu r} \right){e^{ - \mu r}}}}{{{r^3}}}.
\end{equation}
We can rewrite $\chi _1( b)$ as
\begin{equation}\label{eA2}
  \chi _1(b) = - \frac{{gb}}{{8{m^2}}}\int\limits_{-\infty }^\infty  {\frac{{\left( {1 + \mu r} \right){e^{ - \mu r}}}}{{{r^3}}}dz'}.
\end{equation}\label{A3}
On the other hand, if we employ the MacDonald function of zeroth order \cite{45}
\begin{equation}
  K_0(\mu b) = \frac{1}{{2\pi }}\int\limits_0^\infty  {\frac{{{e^{ - \mu r}}}}{r}dz'}  = \frac{1}{{2\pi }}\int\limits_0^\infty  {\frac{{{e^{ - \mu \sqrt {{b^2} + z{'^2}} }}}}{{\sqrt {{b^2} + z{'^2}} }}dz'}
\end{equation}
with the following property
\begin{equation}\label{A4}
  \begin{split}
\frac{d}{db}(K_0(\mu b))=& \frac{1}{2\pi}\frac{d}{db}\int\limits_0^\infty\frac{e^{-\mu r}}{r}dz'= \frac{1}{2\pi}\int\limits_0^\infty \frac{d}{dr}\left(\frac{e^{- \mu r}}{r}\right).\frac{dr}{db}dz'\\
=&-\frac{1}{2\pi}\int\limits_0^\infty\left(\frac{e^{- \mu r}+ \mu re^{- \mu r}}{r^2}\right).\frac{b}{r}dz'= - \frac{b}{2\pi}\int\limits_0^\infty \left(\frac{1 + \mu r}{r^3}\right)e^{- \mu r}dz'.
  \end{split}
\end{equation}
one gets
\begin{equation}\label{A5}
\chi_1(b)= \frac{\pi g}{4m^2}\frac{d}{{db}}\left(K_0(\mu b)\right) =  - \frac{\mu \pi g}{4m^2}K_1(\mu b),
\end{equation}
where $K_1(\mu b)$ is the MacDonald function of first order,$K_1(\mu b)=-\frac{1}{\mu}\frac{d}{db}\left( K_0(\mu b)\right)$.\\
Now, we turn to the calculation of $\chi _0(b)$
\begin{equation}\label{A6}
{\chi _0}\left( b \right) = \frac{1}{{2ip}}\int\limits_{ - \infty }^\infty  {\left[ {U\left( r \right) + \frac{1}{{8{m^2}}}\left[ {{\nabla ^2}U\left( r \right)} \right]} \right]} dz'.
\end{equation}
For the Yukawa potential (\ref{e41}), we get
\begin{equation}\label{A7}
\nabla ^2U\left( r \right) = \frac{1}{{{r^2}}}\frac{\partial }{{\partial r}}\left( {{r^2}\frac{{\partial U}}{{\partial r}}} \right) =  - \frac{g}{{{r^2}}}\frac{\partial }{{\partial r}}\left[ {\left( {1 + \mu r} \right){e^{ - \mu r}}} \right] = \frac{{{\mu ^2}g{e^{ - \mu r}}}}{r}
\end{equation}
Thus
\begin{equation}\label{A8}
\chi_0(b)= \frac{g}{{2ip}}\left( {1 + \frac{{{\mu ^2}}}{{8{m^2}}}} \right)\int\limits_0^\infty  {\frac{{{e^{ - \mu r}}}}{r}} dz' = \frac{{\pi g}}{{ip}}\left( {1 + \frac{{{\mu ^2}}}{{8{m^2}}}} \right){K_0}\left( {\mu b} \right)
\end{equation}
Substitution of (\ref{A5}) and (\ref{A8}) into (\ref{e320}) and (\ref{e321}) one gets
\begin{equation}\label{A9}
  \begin{split}
A(\theta)=&- ip\int\limits_0^\infty  {bdb} {J_0}(b\Delta )\left\{ {\exp \left( {\frac{{\pi g}}{{ip}}\left( {1 + \frac{{{\mu ^2}}}{{8{m^2}}}} \right){K_0}(\mu b)} \right).{\rm{cos}}\left( { - \frac{{\mu \pi g}}{{8{m^2}}}{K_1}\left( {\mu b} \right)} \right) - 1} \right\}\\
 \simeq &- \pi g\left( {1 + \frac{{{\mu ^2}}}{{8{m^2}}}} \right)\int\limits_0^\infty  {bdb} {J_0}(b\Delta ){K_0}(\mu b) = \frac{{ - \pi g\left( {1 + \frac{{{\mu ^2}}}{{8{m^2}}}} \right)}}{{{\Delta ^2} + {\mu ^2}}}
\end{split}
\end{equation}
\begin{equation}\label{A10}
  \begin{split}
B(\theta)=&- ip\int\limits_0^\infty  {bdb} {J_1}(b\Delta )\exp \left( {\frac{{\pi g}}{{ip}}\left( {1 + \frac{{{\mu ^2}}}{{8{m^2}}}} \right){K_0}(\mu b)} \right){\rm{sin}}\left( { - \frac{{\mu \pi g}}{{4{m^2}}}{K_1}\left( {\mu b} \right)} \right)\\
\simeq &ip\int\limits_0^\infty  {bdb} {J_1}(b\Delta )\left[ {1 + \frac{{\pi g}}{{ip}}\left( {1 + \frac{{{\mu ^2}}}{{8{m^2}}}} \right){K_0}(\mu b)} \right]\left[ {\frac{{\mu \pi g}}{{4{m^2}}}{K_1}\left( {\mu b} \right)} \right]\\
 \simeq & \frac{{i\mu \pi gp}}{{4{m^2}}}\int\limits_0^\infty  {bdb} {J_1}(b\Delta ){K_1}\left( {\mu b} \right) = \frac{{i\pi gp}}{{4{m^2}}}.\frac{\Delta }{{{\mu ^2} + {\Delta ^2}}}.
  \end{split}
\end{equation}
The differential cross section is then
\begin{equation}\label{A11}
  \begin{split}
\frac{{d\sigma }}{{d\Omega }}\left| {_{{Y_D}}}\right. =& {\left| {A\left( \theta  \right)} \right|^2} + {\left| {B\left( \theta  \right)} \right|^2} = \frac{{{\pi ^2}{g^2}}}{{{{\left( {{\Delta ^2} + {\mu ^2}} \right)}^2}}}\left[ {{{\left( {1 + \frac{{{\mu ^2}}}{{8{m^2}}}} \right)}^2} + \frac{{{p^2}{\Delta ^2}}}{{16{m^4}}}} \right]\\
 =& \frac{{{\pi ^2}{g^2}}}{{{{\left( {4{p^2}{{\sin }^2}\left( {\frac{\theta }{2}} \right) + {\mu ^2}} \right)}^2}}}\left[ {{{\left( {1 + \frac{{{\mu ^2}}}{{8{m^2}}}} \right)}^2} + \frac{{{p^4}{{\sin }^2}\left( {\frac{\theta }{2}} \right)}}{{4{m^4}}}} \right].
\end{split}
\end{equation}
This expression of differential cross section is for the case in which the Darwin term is included. If we ignore this term in (\ref{e46}), the differential cross section is
\begin{equation}\label{A12}
\frac{{d\sigma }}{{d\Omega }}\left| {_{{Y_o}}} \right. = \frac{{{\pi ^2}{g^2}}}{{{{\left( {{\mu ^2} + {\Delta ^2}} \right)}^2}}}\left( {1 + \frac{{{p^2}{\Delta ^2}}}{{16{m^4}}}} \right) = \frac{{{\pi ^2}{g^2}}}{{{{\left( {{\mu ^2} + 4{p^2}{{\sin }^2}(\theta /2)} \right)}^2}}}\left( {1 + \frac{{{p^4}{{\sin }^2}(\theta /2)}}{{4{m^4}}}} \right).
\end{equation}
\section{{Appendix B. Gauss Potential}}
For this case, we do similar to for the case of Yukawa potential. In particular, we have
\begin{align}
{\nabla ^2}U = \frac{1}{{{r^2}}}\frac{\partial }{{\partial r}}\left( {{r^2}\frac{{\partial U}}{{\partial r}}} \right) = & - \frac{{2\lambda \alpha }}{{{r^2}}}\frac{\partial }{{\partial r}}\left( {{r^3}{e^{ - \alpha {r^2}}}} \right) =  - 2\lambda \alpha \left( {3 - 2\alpha {r^2}} \right){e^{ - \alpha {r^2}}}\label{B1}\\
\left[ {U\left( r \right) + \frac{1}{{8{m^2}}}\left[ {{\nabla ^2}U\left( r \right)} \right]} \right] =& \lambda \left( {1 - \frac{{3\alpha }}{{4{m^2}}}} \right){e^{ - \alpha {r^2}}} + \frac{{\lambda {\alpha ^2}}}{{2{m^2}}}{r^2}{e^{ - \alpha {r^2}}}\label{B2}
\end{align}
Substitution of (\ref{B2}) into (\ref{e314}) and (\ref{e315}) yields the final expressions for $\chi_0(b,\infty)$ and $\chi_1(b,\infty)$
\begin{equation}\label{B3}
\begin{split}
\chi _0(b) =& \frac{1}{{2ip}}\int\limits_{ - \infty }^\infty  {\left[ {\lambda \left( {1 - \frac{{3\alpha }}{{4{m^2}}}} \right){e^{ - \alpha {r^2}}} + \frac{{\lambda {\alpha ^2}}}{{2{m^2}}}{r^2}{e^{ - \alpha {r^2}}}} \right]} dz'\\
 =& \frac{\lambda }{{2ip}}\left( {1 - \frac{{3\alpha }}{{4{m^2}}}} \right)\int\limits_{ - \infty }^\infty  {{e^{ - \alpha \left( {{b^2} + z{'^2}} \right)}}dz'}  + \frac{{\lambda {\alpha ^2}}}{{4ip{m^2}}}\int\limits_{ - \infty }^\infty  {\left( {{b^2} + z{'^2}} \right){e^{ - \alpha \left( {{b^2} + z{'^2}} \right)}}dz'} \\
 =& \frac{\lambda }{{2ip}}\left( {1 - \frac{{3\alpha }}{{4{m^2}}} + \frac{{{\alpha ^2}{b^2}}}{{2{m^2}}}} \right)\exp \left( { - \alpha {b^2}} \right)\sqrt {\frac{\pi }{\alpha }}  + \frac{{\lambda \alpha }}{{8ip{m^2}}}\exp \left( { - \alpha {b^2}} \right)\sqrt {\frac{\pi }{\alpha }}\\
 =& \frac{\lambda }{{2ip}}\left( {1 - \frac{\alpha }{{2{m^2}}} + \frac{{{\alpha ^2}{b^2}}}{{2{m^2}}}} \right)\exp \left( { - \alpha {b^2}} \right)\sqrt {\frac{\pi }{\alpha }} \\
 =&{C_1}\exp \left( { - \alpha {b^2}} \right) + {C_2}{b^2}\exp \left( { - \alpha {b^2}} \right)
 \end{split}
\end{equation}
\begin{equation}\label{B4}
  \begin{split}
\chi_1(b)=& \frac{b}{{8{m^2}}}\int\limits_{ - \infty }^\infty  {\frac{1}{r}\frac{{dU}}{{dr}}dz'}= - \frac{{\alpha b\lambda }}{{4{m^2}}}\int\limits_{ - \infty }^{ + \infty } {{e^{ - \alpha \left( {{b^2} + z{'^2}} \right)}}dz'} \\
 =&- \frac{{\alpha b\lambda {e^{ - \alpha {b^2}}}}}{{4{m^2}}}\sqrt {\frac{\pi }{\alpha }}  =  - \frac{{\sqrt {\alpha \pi } \lambda b{e^{ - \alpha {b^2}}}}}{{4{m^2}}} = {C_3}b{e^{ - \alpha {b^2}}},
  \end{split}
\end{equation}
where
\begin{equation}\label{B5}
{C_1} = \sqrt {\frac{\pi }{\alpha }} \frac{\lambda }{{2ip}}\left( {1 - \frac{\alpha }{{2{m^2}}}} \right);{C_2} = \sqrt {\frac{\pi }{\alpha }} \frac{{\lambda {\alpha ^2}}}{{4ip{m^2}}};{C_3} =  - \frac{{\sqrt {\alpha \pi } \lambda }}{{4{m^2}}}.
\end{equation}
Substituting the expressions (\ref{B3}) and (\ref{B4}) into (\ref{e320}) and (\ref{e321}), performing  the Taylor’s approximation, and keeping only the terms up to the first order, one gets
\begin{equation}\label{B6}
  \begin{split}
A(\theta )=& - ip\int\limits_0^\infty  {bdb} {J_0}(b\Delta )\left[ {\exp \left( {{C_1}\exp \left( { - \alpha {b^2}} \right) + {C_2}{b^2}\exp \left( { - \alpha {b^2}} \right)} \right)c{\rm{os}}\left( {{C_3}{e^{ - \alpha {b^2}}}} \right) - 1} \right]\\
 \simeq & - ip\int\limits_0^\infty  {bdb} {J_0}(b\Delta )\left\{ {{C_1} + {C_2}{b^2}} \right\}{e^{ - \alpha {b^2}}}\\
 =& - i{C_1}p\frac{{\exp \left( { - \frac{{{\Delta ^2}}}{{8\alpha }}} \right)}}{{\Delta \sqrt \alpha  }}{M_{\frac{1}{2},0}}\left( {\frac{{{\Delta ^2}}}{{4\alpha }}} \right) - i{C_2}p\frac{{\exp \left( { - \frac{{{\Delta ^2}}}{{8\alpha }}} \right)}}{{\Delta {\alpha ^{\frac{3}{2}}}}}{M_{\frac{3}{2},0}}\left( {\frac{{{\Delta ^2}}}{{4\alpha }}} \right)
 \end{split}
\end{equation}
and
\begin{equation}\label{B7}
  \begin{split}
B(\theta ) = &ip\int\limits_0^\infty  {bdb} {J_1}(b\Delta )\exp \left( {{C_1}\exp \left( { - \alpha {b^2}} \right) + {C_2}{b^2}\exp \left( { - \alpha {b^2}} \right)} \right){\rm{sin}}\left( {{C_3}b{e^{ - \alpha {b^2}}}} \right)\\
 \simeq &i{C_3}p\int\limits_0^\infty  {{b^2}db} {J_1}(b\Delta ){e^{ - \alpha {b^2}}} = i{C_3}p\frac{{\exp \left( { - \frac{{{\Delta ^2}}}{{8\alpha }}} \right)}}{{\alpha \Delta }}{M_{1,\frac{1}{2}}}\left( {\frac{{{\Delta ^2}}}{{4\alpha }}} \right)
 \end{split}
\end{equation}
In (\ref{B6}) and (\ref{B7}), the following integral identity for Bessel functions has been used
\begin{equation}\label{B8}
\int\limits_0^\infty  {{x^\mu }{e^{ - \alpha {x^2}}}{J_v}\left( {\beta x} \right)dx = \frac{{\Gamma \left( {\frac{1}{2}\mu  + \frac{1}{2}\nu  + 1} \right)}}{{\beta {\alpha ^{\frac{1}{2}\mu }}\Gamma \left( {\nu  + 1} \right)}}\exp \left( { - \frac{{{\beta ^2}}}{{8\alpha }}} \right)} {M_{\frac{1}{2}\mu ,\frac{1}{2}\nu }}\left( {\frac{{{\beta ^2}}}{{4\alpha }}} \right),
\end{equation}
where $Re(\alpha)> 0;Re(\mu + \nu)> - 1$ and $M_{\mu,\nu}(z)$  is the Whittaker function.\\
With the notice that
\begin{equation}\label{B9}
  \begin{split}
M_{\mu ,\nu}\left( {\frac{{{\Delta ^2}}}{{4\alpha }}} \right) =& \exp \left( { - \frac{{{\Delta ^2}}}{{8\alpha }}} \right).{\left( {\frac{{{\Delta ^2}}}{{4\alpha }}} \right)^{\nu  + \frac{1}{2}}}.{}_1{F_1}\left( {\nu  - \mu  + \frac{1}{2};1 + 2\nu ;\frac{{{\Delta ^2}}}{{4\alpha }}} \right)\\
 =& \exp \left( { - \frac{{{\Delta ^2}}}{{8\alpha }}} \right).{\left( {\frac{{{\Delta ^2}}}{{4\alpha }}} \right)^{\nu  + \frac{1}{2}}}\sum\limits_{n = 0}^\infty  {\frac{{{a^{(n)}}}}{{{b^{(n)}}n!}}} {\left( {\frac{{{\Delta ^2}}}{{4\alpha }}} \right)^n},
\end{split}
\end{equation}
where
\begin{equation}\label{B10}
  \begin{split}
  a =& \nu  - \mu  + \frac{1}{2};b = 1 + 2\nu \\
{a^{(0)}} =& 1;\quad {a^{(n)}} = a\left( {a + 1} \right)\left( {a + 2} \right) + ...(a + n - 1)
\end{split}
\end{equation}
From (\ref{B9}), one gets
\begin{equation}\label{B11}
  \begin{split}
{M_{\frac{1}{2},0}}\left( {\frac{{{\Delta ^2}}}{{4\alpha }}} \right) =& \exp \left( { - \frac{{{\Delta ^2}}}{{8\alpha }}} \right){\left( {\frac{{{\Delta ^2}}}{{4\alpha }}} \right)^{\frac{1}{2}}}\\
{M_{\frac{3}{2},0}}\left( {\frac{{{\Delta ^2}}}{{4\alpha }}} \right) = &\exp \left( { - \frac{{{\Delta ^2}}}{{8\alpha }}} \right){\left( {\frac{{{\Delta ^2}}}{{4\alpha }}} \right)^{\frac{1}{2}}}\left[ {1 - \frac{{{\Delta ^2}}}{{4\alpha }}} \right]\\
{M_{1,\frac{1}{2}}}\left( {\frac{{{\Delta ^2}}}{{4\alpha }}} \right) =& \exp \left( { - \frac{{{\Delta ^2}}}{{8\alpha }}} \right)\left( {\frac{{{\Delta ^2}}}{{4\alpha }}} \right)
\end{split}
\end{equation}
Substituting (\ref{B11}) into (\ref{B6}) and (\ref{B7}) we obtain the following expressions for $A(\theta)$ and $B(\theta)$
\begin{align}
A(\theta) =&\frac{{\lambda \left( {{\Delta ^2} - 8{m^2}} \right)}}{{32{m^2}\alpha }}\sqrt {\frac{\pi }{\alpha }} \exp \left( { - \frac{{{\Delta ^2}}}{{4\alpha }}} \right),\label{B12}\\
B\left( \theta  \right) =&- i\frac{{\lambda p\Delta }}{{16\alpha {m^2}}}\sqrt {\frac{\pi }{\alpha }} \exp \left( { - \frac{{{\Delta ^2}}}{{4\alpha }}} \right).\label{B13}
\end{align}
The differential cross section is then
\begin{equation}\label{B14}
  \begin{split}
\frac{{d\sigma }}{{d\Omega }}\left| {_{{G_D}}} \right. =& {\left| {A\left( \theta  \right)} \right|^2} + {\left| {B\left( \theta  \right)} \right|^2}\\
 = &\frac{{\pi {\lambda ^2}}}{{16{\alpha ^3}}}\exp \left( { - \frac{{2{p^2}{{\sin }^2}(\theta /2)}}{\alpha }} \right)\left[ {{{\left( {1 - \frac{{{p^2}}}{{2{m^2}}}{{\sin }^2}(\theta /2)} \right)}^2} + \frac{{{p^4}}}{{4{m^4}}}{{\sin }^2}(\theta /2)} \right]
  \end{split}
\end{equation}
Now, if the Darwin term is ignored, the differential cross section is
\begin{equation}\label{B15}
\frac{{d\sigma }}{{d\Omega }}\left| {_{{G_o}}} \right. = \frac{{\pi {\lambda ^2}}}{{16{\alpha ^3}}}\exp \left( { - \frac{{2{p^2}{{\sin }^2}(\theta /2)}}{\alpha }} \right)\left( {1 + \frac{{{p^4}{{\sin }^2}(\theta /2)}}{{4{m^4}}}} \right).
\end{equation}

\end{document}